\documentclass[aps,twocolumn,prb]{revtex4}
\usepackage{epsfig}
\usepackage{epstopdf}
\usepackage{amsmath,amsfonts,paralist}
\usepackage{amssymb}
\usepackage{graphicx}
\usepackage{mathrsfs} 
\usepackage{color}
\usepackage{natbib,hyperref,url}
\usepackage{dcolumn}
\usepackage{bm}
\usepackage{dsfont}
\usepackage{upgreek}

\definecolor{orange}{rgb}{0.9,0.7,0}

\newcommand{\bG     }{\mbox{\boldmath$G$}}
\newcommand{\bphi     }{\mbox{\boldmath$\phi$}}
\newcommand{\bpsi     }{\mbox{\boldmath$\psi$}}

\begin{document}

\title{Transition between localized and extended states in the hierarchical Anderson model}

\author{F. L. Metz$^{1}$, L. Leuzzi$^{1,2}$, G. Parisi$^{1,2,3}$, V. Sacksteder IV$^{4}$}
\affiliation{$^1$ Dip. Fisica, Universit{\`a} {\em La Sapienza}, Piazzale
  A. Moro 2, I-00185, Rome, Italy \\ $^2$ IPCF-CNR, UOS Roma {\em
    Kerberos}, Universit\`a {\em La Sapienza}, P. le A. Moro 2, I-00185,
  Rome, Italy \\ $^3$ INFN, Piazzale A. Moro 2, 00185, Rome, Italy \\$^4$ Institute of Physics, Chinese Academy of Sciences, Beijing 100190, China}
\date{\today}  
\begin{abstract}

We present strong numerical evidence for the existence of a localization-delocalization 
transition in the eigenstates of the 1-D Anderson model with long-range hierarchical hopping.  Hierarchical 
models are important because of the well-known mapping between their phases and those of  models  with short 
range hopping in higher dimensions, and also because  the renormalization group can be applied exactly without the  
approximations that generally are  required in other models.  In the hierarchical Anderson model we find a finite 
critical disorder strength $W_c$ where the average inverse participation ratio goes to zero; at small disorder $W < W_c$ the model 
lies in a delocalized phase.   This result is based on numerical calculation of the inverse participation ratio in the infinite 
volume limit using  an exact renormalization group approach facilitated by the model's hierarchical structure. Our results are 
consistent with the presence of an Anderson transition in short-range models with $D > 2$ dimensions, which was  predicted  
using  renormalization group arguments.  Our finding should 
stimulate  interest in the hierarchical Anderson model as a simplified and tractable model of  the Anderson localization transition 
which occurs in finite-dimensional systems with short-range hopping.

\end{abstract}
\maketitle

\section{Introduction}
After more than fifty years the Anderson transition \cite{Anderson58}
between localized and extended wave-functions
of a single quantum particle moving in a disordered medium remains
the focus of considerable interest. \cite{Kramer93,Evers2008} Crucial contributions to this field have been made by  
exactly solvable tight-binding models, such as  
1-D models with nearest-neighbour
hopping \cite{Kramer93,Economoubook1} and models on the Bethe lattice. 
\cite{AAT73,AbouChacra74}
Here we consider another interesting class of tight-binding models  with  long-range hopping
 arranged in a hierarchical block
structure and  decaying according  to a power law with exponent $\alpha$. 
Hierarchical
models 
 have a long history in statistical
physics starting with  
Dyson \cite{Dyson1969}, and (as we will explain later) they provide an indirect route to understanding 
phases and critical behaviour in $D$-dimensional systems. \cite{Molchanov1996}

We study the hierarchical Anderson model (HAM) introduced by Bovier, 
which combines  on-site disorder with hierarchically-structured long range  hopping. \cite{Bovier}  In the absence of disorder, the spectrum is an infinite set of  highly degenerate flat bands  that accumulate at the upper spectral edge.  The degeneracies are arranged in a geometric series: one half of the pure HAM's states lie in the lowest energy band, one quarter in the next highest energy, etc. 
Hierarchical models 
preserve their structure under renormalization
group transformations, \cite{Baker72,Bovier,Meurice2007}
which has allowed proof of 
 several rigorous results 
 about the site disordered
HAM's spectrum, \cite{Molchanov1996,Kr07,Kritch07,Kr08,Muller2012} and 
may promise exact extensions of the successful scaling theory of localization. \cite{Abrahams79}
In particular, the absolutely continuous part of the spectrum vanishes and
the model presents only spectral localization, provided that the hopping
 decays sufficiently quickly with distance, i.e., the hopping decay exponent $ \alpha > 3/2$. \cite{Kr07,Kritch07}

Unfortunately, much less is known about
the size  of the HAM's eigenvectors.
 The degeneracies of the pure model permit different choices of mutually orthogonal sets of eigenvectors.  The most extended set  
  consists of  infinitely extended plane waves, while the least extended set 
   has sizes that are strongly band-dependent, with very localized states in the lowest band and infinitely extended states in the highest band.  
In the presence of on-site disorder, it recently has been  argued that all states are always localized, \cite{MonthusGarel} based on an analogy
with the criticality results for random-matrix models, such as ensembles of ultrametric \cite{Ossipov2009,bogomolny2011} and power-law random  banded matrices. \cite{Mirlin96} Both models, characterized by an exponent $\alpha$ controlling the power-law decaying 
{\it random} hoppings,  exhibit an
extended phase for $\alpha < 1$ and a localized phase
for $\alpha >1$.
This would rule out
the possibility of a transition in the HAM, which has a well-defined macroscopic
limit only for $\alpha > 1$. 
However models with random hopping are relatively simple: the scattering length vanishes and only the localization length is important. The HAM belongs instead to the  class of models with {\it deterministic} hopping, which are much richer because they have non-trivial physics at both  length scales. 
In particular, the 1-D Anderson model with on-site disorder and deterministic power-law hopping exhibits a localization transition at its upper spectral edge.
 \cite{Yeung87,Rodriguez2000,Rodriguez2003,Malyshev2004,Moura2005}
%

In this work we show that the  
HAM exhibits a localization-delocalization transition near its upper spectral edge.  We perform
a thorough numerical study of the inverse participation ratio, which is the inverse of the eigenstate volume.       Thanks to the HAM's invariance under  block renormalization group (RG) transformations,  we obtain recurrence equations for calculating the resolvent matrix. This recursive method allows us  to calculate the IPR  in 
systems large enough to  precisely determine their infinite size behavior.
Our results also suggest  that there is a critical value of $\alpha$ above which all states are localized, in 
analogy with the lower critical dimension $D = 2$ below which finite-dimensional
short-range systems are always localized and above which an Anderson transition was predicted using RG arguments. \cite{Abrahams79}


%
\begin{figure}[t!]
\center
\begin{picture}(300,100)(12,-10)
  \put(40,0.28){\circle*{4.5}}  
  \put(70,0.28){\circle*{4.5}}  
  \put(100,0.28){\circle*{4.5}}  
  \put(130,0.28){\circle*{4.5}}  
  \put(160,0.28){\circle*{4.5}}  
  \put(190,0.28){\circle*{4.5}}  
  \put(220,0.28){\circle*{4.5}}  
  \put(250,0.28){\circle*{4.5}}

  \put(37,-11){{\small $\epsilon_{1}$}}
  \put(67,-11){{\small $\epsilon_{2}$}}
  \put(97,-11){{\small $\epsilon_{3}$}}
  \put(127,-11){{\small $\epsilon_{4}$}}
  \put(157,-11){{\small $\epsilon_{5}$}}
  \put(187,-11){{\small $\epsilon_{6}$}}
  \put(217,-11){{\small $\epsilon_{7}$}}
  \put(247,-11){{\small $\epsilon_{8}$}}

  \put(40,0.28){\line(0, 1){20.0}}
  \put(70,0.28){\line(0, 1){20.0}}
  \put(100,0.28){\line(0, 1){20.0}}
  \put(130,0.28){\line(0, 1){20.0}}
  \put(160,0.28){\line(0, 1){20.0}}
  \put(190,0.28){\line(0, 1){20.0}}
  \put(220,0.28){\line(0, 1){20.0}}
  \put(250,0.28){\line(0, 1){20.0}}

  \put(40,20.28){\line(1, 0){30.0}}
  \put(100,20.28){\line(1, 0){30.0}}
  \put(160,20.28){\line(1, 0){30.0}}
  \put(220,20.28){\line(1, 0){30.0}}

  \put(55,20.28){\line(0, 1){20.0}}
  \put(115,20.28){\line(0, 1){20.0}}
  \put(175,20.28){\line(0, 1){20.0}}
  \put(235,20.28){\line(0, 1){20.0}}

  \put(55,40.28){\line(1, 0){60.0}}
  \put(175,40.28){\line(1, 0){60.0}}

  \put(85,40.28){\line(0, 1){20.0}}
  \put(205,40.28){\line(0, 1){20.0}}

  \put(85,60.28){\line(1, 0){120.0}}

  \put(53,10){{\footnotesize $t_1$}}
  \put(113,10){{\footnotesize $t_1$}}
  \put(173,10){{\footnotesize $t_1$}}
  \put(233,10){{\footnotesize $t_1$}}

  \put(83,30){{\footnotesize $t_2$}}
  \put(203,30){{\footnotesize $t_2$}}

  \put(143,50){{\footnotesize $t_3$}}

  \put(14,17.0){{\footnotesize $p=1$}}
  \put(14,37.0){{\footnotesize $p=2$}}
  \put(14,57.0){{\footnotesize $p=3$}}

\end{picture}
\caption{Schematic representation of the hierarchical Anderson 
model, cf. Eq. (\ref{hamilt}), with $L=2^{3}$ sites.
Lines denote  hopping energies $t_p$ between sites in
 distinct blocks of size $2^{p-1}$. }
\label{picthierarc}
\end{figure}
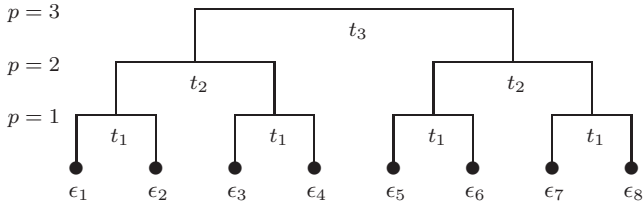

\section{The hierarchical Anderson model.}
The HAM is a 1-D tight binding model
with $L=2^{N}$ equally spaced sites,  and independently distributed random site potentials $\epsilon_{i}$,
   $i=1,\dots,L$. 
The Hamiltonian reads

\begin{eqnarray}
\mathcal{H}_{N} &=& \sum_{i=1}^{2^{N}} \epsilon_{i} \mid i \rangle \langle i \mid
\label{hamilt}
\\
 &+& \sum_{p=1}^{N} V_{p} \sum_{r=1}^{2^{N-p}} 
\sum_{i\neq j}^{1,2^{p}}  \mid (r-1)2^p + i \rangle  \langle (r-1)2^p + j \mid \,,
\nonumber
\end{eqnarray}
where $| i \rangle $
is the canonical site basis.  The second line is the hierarchical hopping matrix introduced by Dyson. \cite{Dyson1969}  It is the heart of the hierarchical Anderson model, and is organized in a tree as illustrated in Fig. \ref{picthierarc}.  The highest level of the tree has index $p = N$ and the lowest level has index $p = 1$.  At each level the system is divided into $2^{N-p}$ separate blocks, each of which contains $2^p$ sites.  The hopping between any two sites within a single block has energy $V_p$.  As seen in Fig. \ref{picthierarc}, the hopping between sites in two different blocks is determined by levels higher in the hierarchy and has energy
$t_p = \sum_{n=p}^{N} V_n$.
We study the deterministic HAM, which has hopping energies   $ V_{p}  = 2^{-\alpha(p-1)}$.
This exponential  decay in the level index $p$ ensures that in large $N \gg 1$ systems the hopping energy between sites separated by a distance $O(L)$ decays according to a power law $t_p \propto O(L^{-\alpha})$, the same as 1-D Anderson models with power-law  hopping. 
\cite{Yeung87,Rodriguez2000,Rodriguez2003,Malyshev2004,Moura2005}
\\\indent 
We study the infinite volume limit of  the  average density of states (DOS) $\rho(E)$ and of the inverse participation ratio $P(E)$.  The former is defined as 
\begin{equation}
\rho(E) = \lim_{L \rightarrow \infty} \Big{\langle} \frac{1}{L} \sum_{\mu=1}^{L} \delta(E - E_{\mu}) \Big{\rangle} \,,
\label{DOSdef}
\end{equation}
where $\langle \dots \rangle$ is the  average with
respect to the disorder potential $ \epsilon_{i} $ and
 $E_{\mu} $ 
  are the HAM's eigenvalues.  
 The DOS measures  the averaged spectrum, but does not contain any signal of the eigenstates' localization or delocalization.  We therefore study  the average inverse participation ratio (IPR) of the normalized eigenstates 
$|\psi_\mu \rangle$: \cite{Wegner80,Metz2010,Slanina2012}
\begin{equation}
P(E) = \lim_{L \rightarrow \infty} \frac{1}{L \rho(E)}  \Big{\langle} \sum_{\mu=1}^{L} I^{L}_\mu \delta(E - E_{\mu}) \Big{\rangle}\,,
\label{avIPR}
\end{equation}
where  
$I^{L}_\mu = \sum_{i=1}^{L} (\langle i | \psi_{\mu} \rangle)^{4}$ is the IPR of an individual eigenstate. Its inverse measures 
 the eigenstate's volume. 
The IPR
is restricted to the interval $0 \le P(E) \le 1$.
States that are perfectly localized on a single site satisfy $P(E) = 1$,  and states that are equally distributed across all sites satisfy $P(E) = 1/L \rightarrow 0$.  
\\\indent   
 In the pure $W=0$ HAM  the DOS
is a series of flat bands  $\rho_{{\rm pure}}(E) = \sum_{p=1}^{\infty} 2^{-p} \delta(E - 
E^{{\rm pure}}_{p-1} )$. Each flat band  is related to a level in the HAM's
hierarchy.
 The bands'  degeneracy decreases repeatedly by factors of two as one moves to higher energy, thus yielding
 the factor $2^{-p}$.
The difference between consecutive
energetic levels falls off as $E^{{\rm pure}}_{p+1} - E^{{\rm pure}}_{p} 
\propto 2^{-(\alpha-1)p}$ and, hence, these accumulate
at  the upper spectral edge $E^{{\rm pure}}_{\infty}$. 
Near $E^{{\rm pure}}_{\infty}$ the integrated 
density of states $\mathcal{N}(E^{{\rm pure}}_{p}) = \sum_{\ell=1}^{p} 2^{-\ell}$ 
follows a power law similar to that of short-range finite-dimensional
systems: $\mathcal{N}(E^{{\rm pure}}_{p}) = 1 - C 
\left( E^{{\rm pure}}_{\infty} - E^{{\rm pure}}_{p} \right)^{d_s/2}$. \cite{Molchanov1996, Kr07,Kr08,Muller2012}
Here $d_s = 2/(\alpha - 1)$ is the {\em spectral dimension} which controls both diffusion
and the long-distance physics of second-order phase transitions such as the Anderson transition.
For $W >0$, the integrated DOS of the HAM exhibits
a Lifshitz tail at the upper spectral edge, with a Lifshitz
exponent given by the spectral dimension. \cite{Muller2012}
This is {\em{ the same behavior as observed in short-range finite-dimensional systems with on-site 
disorder}}, where the integrated DOS exhibits a Lifshitz tail controlled
by the Euclidean dimension. \cite{Kirsch83}
Overall, as $E^{{\rm pure}}_{\infty}$  is approached,
the spectral properties of the pure HAM become similar to short-range finite-dimensional systems. 
This is, thus, the most promising region for studying localization transitions. 
Much of the interest in hierarchical models originates in their mapping to short-range models whose 
Euclidean dimension is strictly related to the spectral dimension
(see, e.g., Ref. [\onlinecite{Ibanez12}] and Refs. therein).


\section{Renormalization equations for the resolvent}
We obtain the DOS and IPR from the diagonal elements of the  
 resolvent matrix $\bG^{(N)}(z) = (z - \mathcal{H}_{N})^{-1}$, where   $z = E - i \eta$, and $\eta$ is a small positive regularizer that smooths our numerical results over an interval in the spectrum with width proportional to $\eta$. \cite{Janssen1998,Song2007}
  We use the following formulas: \cite{Economou72,Fyodorov1991,Wegner80,Metz2010}
\begin{eqnarray}
\rho(E) &=&\lim_{\eta \rightarrow 0^{+}} \lim_{L \rightarrow \infty} \frac{1}{L \pi} \sum_{i=1}^{L} \Big{\langle} {\rm Im} \, G_{i}^{(N)}(z) \Big{\rangle} \,,
\label{rhoResolv} 
\\
P(E) &=& \lim_{\eta \rightarrow 0^{+}} \lim_{L \rightarrow \infty} \frac{\eta}{\pi L \rho(E)} 
\sum_{i=1}^{L} \Big{\langle} | G_{i}^{(N)}(z) |^{2} \Big{\rangle} \,.
\label{IPRResolv} 
\end{eqnarray}
%
%
%
The HAM's hierarchical structure allowed us to develop a block RG approach which recursively calculates the resolvent  for one instance of the disorder. 
Our calculation has two phases: a sweep up the hierarchy, and then a sweep back down.  At each step $\ell$  of the sweep up we remove the basis states associated with one flat band, and calculate an energy-dependent effective Hamiltonian which acts in the reduced basis but exhibits the same poles found in the original full-basis Hamiltonian.  This effective Hamiltonian  retains the hierarchical form but  its hopping energies $\{ V_p^{(\ell)} \}$ and disorder potentials
$\{ \mu_i^{(\ell)} \}$ are renormalised according to
\begin{eqnarray}
\mu^{(\ell)}_i  &=& \frac{2 \mu^{(\ell-1)}_{2i-1} 
\mu^{(\ell-1)}_{2i} }{\mu^{(\ell-1)}_{2i-1} + \mu^{(\ell-1)}_{2i} }
+ 2 V^{(\ell-1)}_1, \, i = 1,. . , 2^{N-\ell} \label{muRecur}  \label{VRecur1} \\
V^{(\ell)}_{p} &=& 2 V^{(\ell-1)}_{p+1} \,, \qquad p=1,\dots,N-\ell \,. \label{VRecur2}
\end{eqnarray}
Hopping energies and disorder potentials at the beginning of the sweep up, $V^{(0)}_{p} = V_{p}$ and  $\mu^{(0)}_i  = \epsilon_{i} - z - \sum_{p=1}^{N} V_{p}$, are those of the original hierarchical Hamiltonian. After $\ell = N$ steps we reach the top of the hierarchy and obtain a single site effective Hamiltonian with disorder potential  $\mu^{(N)}_{1}$.  The resolvent 
 of this Hamiltonian is simply $G^{0}_{1}(z) =  - 1/\mu^{(N)}_{1}$.  We use this resolvent to begin  the sweep back down, in which we progressively restore the original basis and recursively calculate the resolvent's diagonal elements in the restored basis:
 \begin{eqnarray}
G_{2i-1}^{(N-\ell+1)}(z) &=& 2 \left[\frac{ \mu^{(\ell-1)}_{2i}}
{\gamma_i^{(\ell-1)}} \right]^2 
G_{i}^{(N-\ell)}(z)- \frac{1}{\gamma_i^{(\ell-1)}} \,,  \label{GreenRecur1}  \\
G_{2i}^{(N-\ell+1)}(z) &=& 2 \left[\frac{ \mu^{(\ell-1)}_{2i-1}}
{\gamma_i^{(\ell-1)}} \right]^2 
G_{i}^{(N-\ell)}(z) - \frac{1}{\gamma_i^{(\ell-1)}} \,,  \label{GreenRecur2}
\end{eqnarray}
with $\gamma_i^{(\ell-1)}= \mu^{(\ell-1)}_{2i-1} + \mu^{(\ell-1)}_{2i} $.  This procedure yields the diagonal elements of the resolvent in the original system, and its memory consumption and computational time grow only linearly with $L$. \footnote{Mathematically Eqs. (\ref{VRecur1}) and (\ref{VRecur2}) are equivalent to recursive 
computation of the Schur complement. }
The derivation of Eqs. (\ref{VRecur1}-\ref{GreenRecur2}) is presented in App. \ref{app:A}.


\begin{figure}[t!]
\center
\includegraphics[scale=1.45]{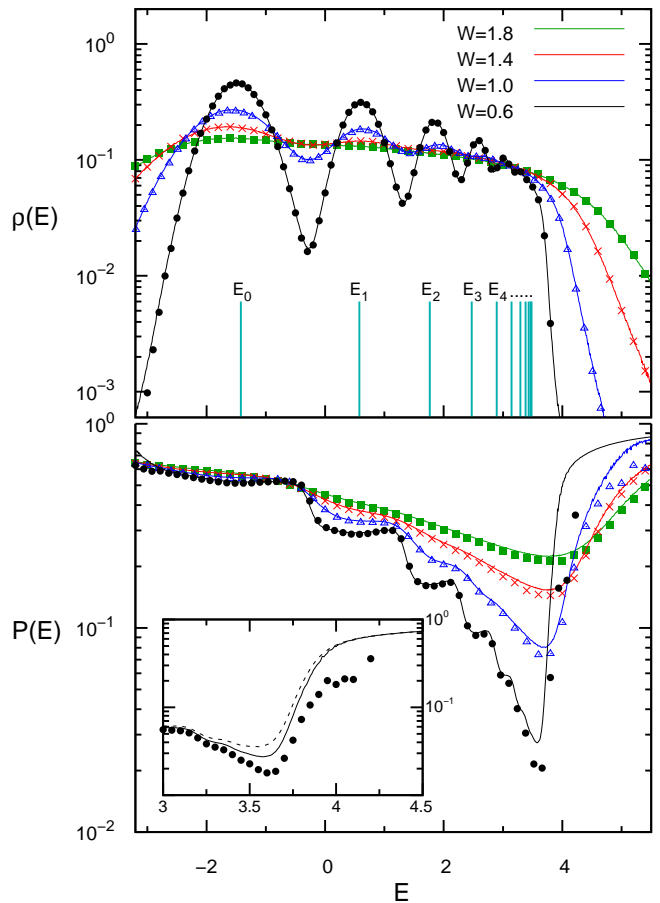}
\caption{
Comparison of the HAM's average DOS and IPR: numerical diagonalization (filled circles) vs.  RG method (solid lines), with hopping decay exponent $\alpha = 7/4$.  The  energies of the pure model's flat bands are marked with cyan vertical lines in the upper pane.  
The inset shows that when the IPR is small the RG method is sensitive to the spectral line width $\eta$; the dashed  and solid lines were obtained   with $\eta=0.01$ and $\eta=0.005$ respectively. }
\label{compDiag}
\end{figure}

\section{Results} 
Fig. \ref{compDiag} compares  the DOS and IPR calculated with  our renormalization  method (solid lines) and $\eta=0.005$ to standard numerical diagonalization (filled circles) in a  system of size $L = 2^{10}$.
The potential $\epsilon_i$ is generated
from a Gaussian distribution 
with zero mean and standard deviation $W$.
Diagonalization results  are averaged over ${\cal N}_\epsilon =  10^{3}$ disorder realizations and renormalisation results over ${\cal N}_\epsilon = 2 \times 10^{4}$ realizations.  Fig. \ref{compDiag} shows excellent agreement between the two methods.

The only important discrepancy  is found in the IPR  at small disorder $W = 0.6$, 
where the DOS falls precipitously.    The observed discrepancy is explained by Fig. \ref{compDiag}'s inset, which compares 
results with two values of the regularization  parameter: $\eta = 0.01$ and $0.005$. The latter lies closer to the diagonalization results, which indicates that when the disorder is small   the limit $\eta \rightarrow 0^{+}$ is reached only at  $\eta \ll 0.005$.

\begin{figure}[t!]
\center
\includegraphics[scale=1.1]{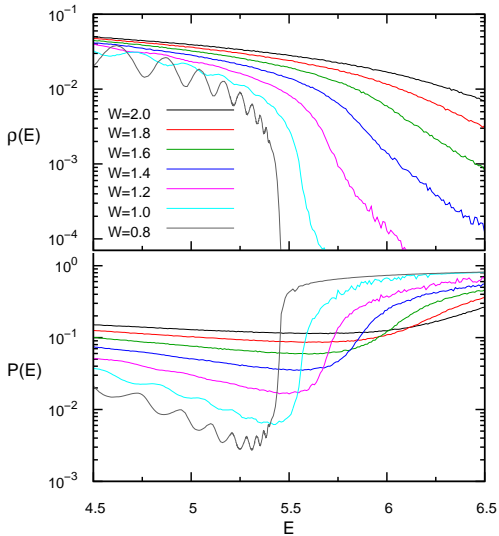}
\caption{
Average DOS and IPR near the upper spectral edge in a large $L=2^{23}$ system at several disorder strengths 
and $\alpha=3/2$, $\eta=5 \times 10^{-4}$ and  ${\cal N}_\epsilon=30$. }
\label{IPRalpha1_5}
\end{figure}

Fig. \ref{compDiag} also gives an overview of the DOS and IPR across the spectrum for a representative hopping decay exponent $\alpha=7/4$ and four different values of the  disorder strength $W = 0.6, \,1.0, \,1.4, \, 1.8$.  At small disorder $W = 0.6$  the average DOS is separated into several bands whose positions coincide with the pure system's flat bands. The associated minima in $P(E)$ show that the eigenstates are bigger in the band centers and smaller at the band edges.  When the disorder is increased  the bands progressively blur together and $P(E)$ steadily increases as the eigenstates become ever more localized.  Fig. \ref{IPRalpha1_5}  shows the same behavior at $\alpha = 3/2$ in systems of size $L = 2^{23}$. 
Reaching such large sizes allows us to explore smaller $\eta$ values and obtain detailed results about many bands near the upper spectral edge. 
Indeed, in order to obtain statistically significant results the spectral line width $\eta$
must 
considerably exceed the mean level spacing $[N \rho(E)]^{-1}$.

\begin{figure}[t!]
\center
\includegraphics[scale=0.95]{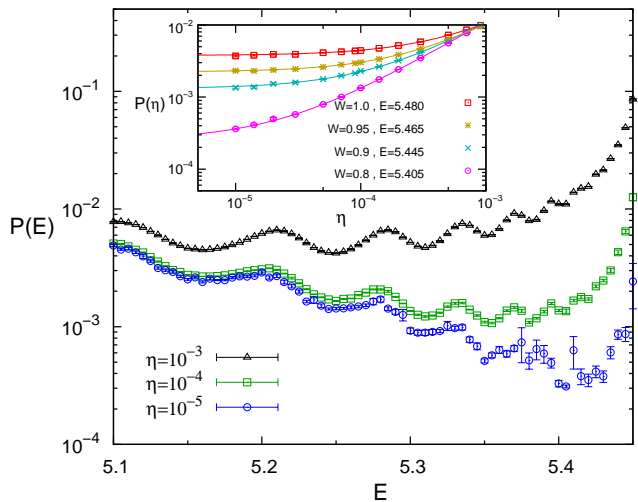}
\caption{The $\eta \rightarrow 0^{+}$ limit.  The three curves show the average IPR
at three values of $\eta$ and $\alpha=3/2$, $W=0.8$, $L=2^{27}$, and ${\cal N}_\epsilon=10$.  The global minimum decreases and shifts to higher energy.   
The inset shows the IPR versus 
$\eta$  at  the  energy of the global minimum.  Again $\alpha=3/2$, but now  $L=2^{28}$ and ${\cal N}_\epsilon=100$.  
The solid lines are linear fits to $P(\eta) = P_{{\rm min}} + b \eta$. 
}
\label{MinIPR}
\end{figure}
In general the IPR exhibits several local minima corresponding to large states 
near the centers of the HAM's bands, and the global  minimum lies near HAM's  upper spectral edge.  
In order to verify the existence of extended
states at finite $W$ we  focus on  the asymptotic value of the global minimum of
 the IPR, $P_{\rm min}(W)$, in the  
  $L \rightarrow \infty$ limit and for infinitesimal  $\eta \rightarrow 0^{+}$.  
The main graph in Fig. \ref{MinIPR} summarizes 
 our calculation of  $P_{\rm min}(W)$ for a particular  hopping decay 
 $\alpha = 3/2$ and disorder strength $W = 0.8$ which lie close to the delocalization transition.  
We display the IPR of a very large $L=2^{27}$ system at three different values of  $\eta$.  
   Statistical errors at smaller $\eta$ are larger because of $\eta$'s  proximity to the level spacing.
   App. \ref{app:B} includes a detailed  discussion of these errors in  the limit $\eta\to 0^{+}$. 
  In particular, we have checked that for $L\geq 2^{26}$ the IPR curves at fixed $\eta$ do not change with $L$,
  which  signals that they accurately represent the infinite volume limit.

The IPR depends on $\eta$, and as $\eta \rightarrow 0^+$ the global minimum deepens and shifts toward higher energy. This effect is not significant at larger disorder $W > 1.0$, but at smaller disorder it forces us to use considerable care with the $\eta\to 0^{+}$ extrapolation. The inset in Fig. \ref{MinIPR} displays our extrapolation  to the limit $\eta\to 0^{+}$ at four weak disorder strengths. 
 At each $W$ we find the energy $E_{\rm min}(W)$  of the local minimum at the lowest value of the spectral width parameter employed, $\eta = 10^{-5}$, and then graph the IPR at that energy as a function of $\eta$.  
  Concerning uncertainty  in $E_{\rm min}(W)$, we have checked that  it affects our results only slightly, and in any case can only cause an unduly careful overestimate of the IPR.  The fitting curves in Fig. \ref{MinIPR} show that the IPR depends linearly on $\eta$ via  $P(\eta) = P_{{\rm min}} + b \eta$.  This allows us to determine very accurately the asymptotic global minimum of the IPR.  

\begin{figure}[t!]
\center
\includegraphics[scale=0.9]{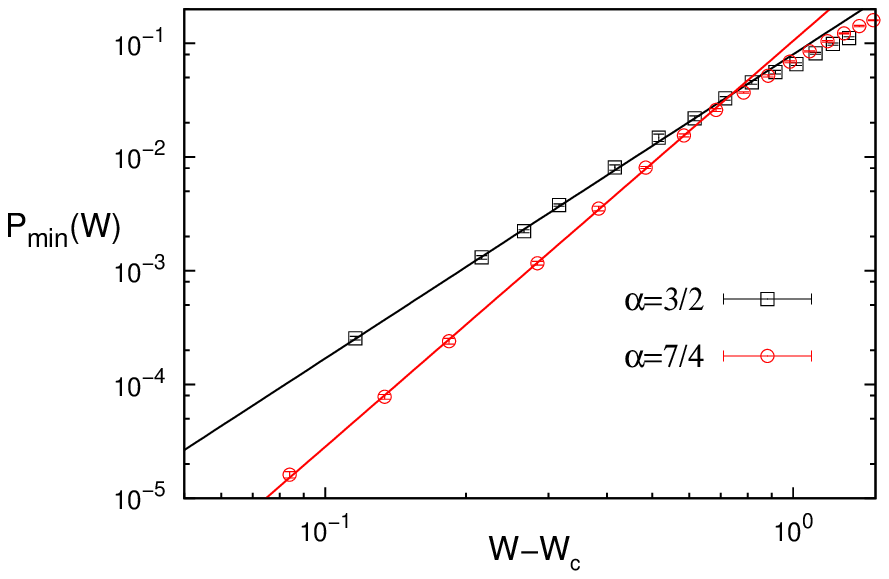}
\caption{Power law behaviour of the minimum IPR $P_{\rm min}(W)$  near the critical disorder strength $W_c$ where it converges to zero.
Solid lines
represent the power-law fit $P_{\rm min}(W)= A (W-W_{c})^{\uppi_{2}}$, with
parameters from
table \ref{tabfitIPRmin}.  
}
\label{MinIPRversusW}
\end{figure}

\begin{table}[t]
\centering
\begin{tabular}{|l|c|c|c|c|c|}
\hline
$\alpha $  & $W_c$ & $\uppi_{2}$ & $A$ & $\chi^{2}/{\rm ndf}$ & ndf \\
\hline
$3/2$ &  $0.684(7)$ & 2.68(7) & 0.080(2) & $0.96$ & $6$ \\
$7/4$  &  $0.016(5)$   & $3.57(8)$ & $0.105(3)$ & $1.21$& $5$ \\

\hline
\end{tabular}
\caption{Values of the parameters and the $\chi^{2}$ of
the power law fit to the IPR data shown in Fig.~\ref{MinIPRversusW}.  $W_c$ is the critical disorder where the delocalization transition occurs, and  $\uppi_2$ is a critical exponent. 
 }
\label{tabfitIPRmin}
\end{table}
The straight lines  in  Fig. \ref{MinIPRversusW}'s log-log plot are the central result of our work: strong numerical evidence that the minimum IPR $P_{\rm min}(W)$ converges to zero according to a power law  $P_{\rm min}(W)= A (W-W_{c})^{\uppi_{2}}$, similar to the power law observed in finite dimensional short-range systems.  \cite{Wegner80,Bauer90,Chang90}
At smaller disorder $W \leq W_c$ the HAM exhibits a delocalized phase.
Table \ref{tabfitIPRmin} reports the best fit parameters for two values of the hopping decay exponent $\alpha = 3/2, 7/4$.    
In both cases our data excludes the possibility that $W_c = 0$. 

 \section{Conclusions.}
 We have analyzed the DOS and the IPR of the   hierarchical Anderson model by means of a RG-based calculation of the resolvent matrix, finding strong  evidence for a localization-delocalization transition at finite disorder  at $\alpha=3/2$ and  $7/4$.  
Since it has been   proven rigorously that the absolutely continuous part of the spectrum vanishes 
 for $\alpha > 3/2$, \cite{Kr07} our results indicate
that  spectral localization  may not imply  the existence of exponentially localized eigenvectors.
A study of the spatial decay of
the resolvent elements should clarify this point and
we expect our work to stimulate further research in this direction.
Our results also 
indicate that the HAM differs from the 1-D tight-binding model with power-law hopping, \cite{Rodriguez2003,Malyshev2004} where  all states are localized 
for $\alpha \ge 3/2$. \cite{Malyshev2004}
Since the HAM's spectral dimension can be mapped to the spatial dimension of Anderson models with short-range hopping, we expect that an Anderson transition exists in the regime
 $1 < \alpha <2$, with $\alpha\simeq 2$ playing a role analogous to the lower critical dimension. 
The presence of extended states in one dimension
is not exclusive to models with long-range hopping, but
it has been also observed in systems with short-range
hopping and correlated on-site disorder. \cite{Phillips90,Lavarda94,Moura98}
Lastly we mention that our RG method can be used to compute  off-diagonal elements of the 
resolvent, allowing determination of  other relevant quantities 
such as the longest localization length. \cite{Kramer93,Economoubook1} 


\acknowledgements
FLM is greatly indebted to Lucas 
Nicolao, Jacopo Rocchi, Pierfrancesco Urbani
and Izaak Neri
for many interesting and useful discussions.
VS thanks Tomi Ohtsuki and Koji Kobayashi for discussions and hospitality.
The research leading to these results has received funding from the European Research Council
(ERC) grant agreement No. 247328 (CriPheRaSy project),
from  the People Programme (Marie Curie Actions) of the European Union's Seventh Framework Programme FP7/2007-2013/ under REA grant agreement No. 290038  (NETADIS project) and 
from the Italian MIUR under the Basic
Research Investigation Fund FIRB2008 program, grant
No. RBFR08M3P4, and under the PRIN2010 program, grant code 2010HXAW77-008.


\appendix

\section{Derivation of the renormalization equations} \label{renormeqs}
\label{app:A}
We discuss here how to derive  the RG equations (6-9) that are used in
the main text.  This calculation can be performed using only linear algebra, but we find it more convenient to use Gaussian integrals. We first 
rewrite $\{ G_{k}^{(N)}(z) \}_{k=1,\dots,L}$, the diagonal elements of the 
resolvent $\bG^{(N)}(z) = (z - \mathcal{H}_{N})^{-1}$,
as the Gaussian integrals 
\begin{eqnarray}
G_{k}^{(N)}(z) &=& \imath  \frac{\int d \bphi  \, \phi_{k}^2
\exp{ \left[     {\cal L}^{(N)}(\phi_{1,\dots,2^{N}}) \right] }}
{\int  d \bphi
\exp{\left[ {\cal L}^{(N)}(\phi_{1,\dots,2^{N}}) \right] } } ,
\label{greenInteg1}
\\
\nonumber
{\cal L}^{(N)}(\phi_{1,\dots,2^{N}})&=& \frac{\imath}{2} \sum_{j=1}^{2^N}  \mu_j \phi_{j}^{2} 
+ W^{(N)}\left(\phi_{1,\dots,2^{N}};V_{1,\dots,N} \right),  
\end{eqnarray}
where $d \bphi = \prod_{j=1}^{2^N}  d \phi_{j}$ and $\mu_{j} =  \epsilon_j -z - \sum_{p=1}^{N} V_{p}$.
The function $W^{(N)}$ encodes the hierarchical hoppings:
\begin{eqnarray}
\nonumber
W^{(N)}\left(  \phi_{1,\dots,2^{N}};V_{1,\dots,N}  \right) \!= \!\frac{\imath}{2} \sum_{p=1}^{N} 
V_{p} \!\sum_{r=1}^{2^{N-p}} \!\!\!\left( \sum_{j=1}^{2^{p}} \phi_{(r-1)2^{p} +j}   \right)^2 .
\end{eqnarray}
We have introduced the simplified notation 
$x_{1,\dots,\mathcal{A}} \equiv x_{1},\dots,x_{\mathcal{A}}$. 
The function ${\cal L}^{(N)}(\phi_{1,\dots,2^{N}})$ has the same form as the HAM's Hamiltonian and therefore  preserves
its formal structure under a RG transformation:  a local
 term incorporating the  random potential and a non-local
hierarchical hopping term.
We make a  change of integration variables
\begin{eqnarray}
\psi^{\pm}_{j} = \frac{1}{\sqrt{2}} (\phi_{2j-1} \pm \phi_{2j} )\,, \qquad
j=1,\dots,2^{N-1}\,, 
\nonumber
\end{eqnarray}
which transforms the hierarchical term
as follows:
\begin{eqnarray}
&&W^{(N)}\left(\phi_{1,\dots,2^{N}};V_{1,\dots,N} \right)
=  \imath V_{1} \sum_{j=1}^{2^{N-1}} (\psi^{+}_{j})^{2} 
\nonumber
\\
&&\qquad\qquad
+ W^{(N-1)}\left(\psi^{+}_{1,\dots,2^{N-1}};V^{\prime}_{1,\dots,N-1} \right),
\nonumber
\end{eqnarray}
where $V^{\prime}_{p} = 2 V_{p}$. 
This transformation allows us to  explicitly calculate the integrals
over $\{ \psi^{-}_{j} \}_{j=1,\dots,2^{N-1}}$ in Eq. (\ref{greenInteg1}), halving the number 
of degrees of freedom. After performing the transformation and integration we obtain 
an equation which relates $\{ G_{i}^{(N)}(z) \}_{i=1,\dots,L}$ for the
original model with $L$ sites to $\{ G_{i}^{(N-1)}(z) \}_{i=1,\dots,L/2}$ for a 
model with $L/2$ sites, but with renormalized parameters.

Partitioning $\{ G_{i}^{(N)}(z) \}_{i=1,\dots,2^{N}}$ into two sectors (one for the even sites and another for the odd sites), we obtain 
the following expressions
\begin{eqnarray}
&&\hspace*{-1.3cm} 
G^{(N)}_{2k-1}(z)
= \frac{\imath}{2} \frac{\int d \bpsi^{+} d \bpsi^{-}  (\psi_{k}^{+} + \psi_{k}^{-})^2
e^{ H^{(N-1)} \left(\psi^{\pm}\right)   }  }
{\int d \bpsi^{+} d \bpsi^{-}
e^{H^{(N-1)} \left(\psi^{\pm}\right)  }  },
\label{Greenodd} \\ 
&&\hspace*{-1.3cm} 
G^{(N)}_{2k}(z)
= \frac{\imath}{2} \frac{\int d \bpsi^{+} d \bpsi^{-}  (\psi_{k}^{+} - \psi_{k}^{-})^2
e^{  H^{(N-1)} \left(\psi^{\pm}\right)  }}
{\int  d \bpsi^{+} d \bpsi^{-} 
e^{   H^{(N-1)} \left(\psi^{\pm} \right)  }}.
\label{Greeneven} 
\end{eqnarray}
In the above expressions we have changed integration variables to
$d \bpsi^{\pm}  = \prod_{j=1}^{2^{N-1}}  d \psi_{j}^{\pm}$ and
 we have defined
\begin{eqnarray} 
&&H^{(N-1)} \left(\psi^{\pm} \right) 
= \frac{\imath}{2} \sum_{j=1}^{2^{N-1}} \sigma_j (\psi_{j}^{+})^2 +
\frac{\imath}{2} \sum_{j=1}^{2^{N-1}} \Delta_j  (\psi_{j}^{-})^2 
\nonumber
\\
&& \hspace*{.5cm}+ \imath \sum_{j=1}^{2^{N-1}} C_j \psi_{j}^{+} \psi_{j}^{-} 
+ W^{(N-1)}\left(\psi^{+}_{1,. .,2^{N-1}};V^{\prime}_{1,. . ,N-1} \right)\,,
\nonumber 
\end{eqnarray}
where the following  quantities are complex valued:
\begin{eqnarray}
\sigma_{j} &=&  \frac{1}{2}\left( \mu_{2j-1} + \mu_{2j} \right) + 2 V_{1},  \nonumber \\ 
\Delta_{j} &=& \frac{1}{2}\left( \mu_{2j-1} + \mu_{2j} \right), \nonumber \\
C_{j} &=& \frac{1}{2} \left( \mu_{2j-1} - \mu_{2j} \right)\,. \nonumber
\end{eqnarray}
Since Eqs. (\ref{Greenodd}) and (\ref{Greeneven}) involve only simple Gaussian
integrals with respect to $\psi^{-}_{1,\dots,2^{N-1}}$, these variables
can be integrated out one by one. 
 We map the resulting expression to   Eq. (\ref{greenInteg1}) for   a system with $2^{N-1}$
sites, renormalized disorder $\mu^{\prime}_{1,\dots,2^{N-1}}$ which obeys Eq. (8) in the main text, and renormalised hopping potential
$V^{\prime}_{1,\dots,N-1}$.  
We obtain 
Eqs. (6-9) at the
first RG step $\ell=1$ of the original model. 
Performing these steps recursively leads
to the
recurrence equations (6) and (7) shown in the main text:
\begin{eqnarray}
G_{2i-1}^{(N-\ell+1)}(z) &=& 2 \left[\frac{ \mu^{(\ell-1)}_{2i}}
{\gamma_i^{(\ell-1)}} \right]^2 
G_{i}^{(N-\ell)}(z)- \frac{1}{\gamma_i^{(\ell-1)}} \,,  \nonumber \\
G_{2i}^{(N-\ell+1)}(z) &=& 2 \left[\frac{ \mu^{(\ell-1)}_{2i-1}}
{\gamma_i^{(\ell-1)}} \right]^2 
G_{i}^{(N-\ell)}(z) - \frac{1}{\gamma_i^{(\ell-1)}} \,,  \nonumber
\end{eqnarray}
where $\gamma_i^{(\ell)} = \mu_{2i-1}^{(\ell)} + \mu_{2i}^{(\ell)}$.


\section{Performing the $\eta \rightarrow 0^{+}$ limit numerically}
\label{app:B}

\begin{figure}[t!]
\includegraphics[scale=0.9]{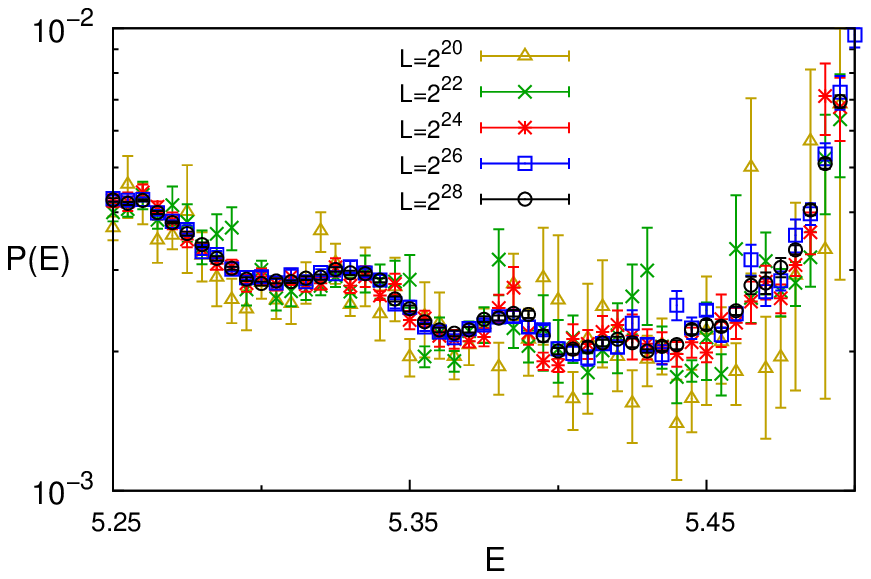}
\caption{Size effects on the IPR at $\alpha=1.5$, $W=0.9$, $\eta=10^{-4}$ and $
{\cal N}_\epsilon=10$. }
\label{sizeeffect}
\end{figure}

The regularization parameter $\eta$ in the resolvent gives each eigenvalue a line width proportional to $\eta$.   This can be
understood by analysing  our equation for the DOS
\begin{equation}
\rho(E) = \lim_{\eta \rightarrow 0^{+}} \lim_{L \rightarrow \infty} \frac{1}{L \pi} \sum_{i=1}^{L} \Big{\langle} {\rm Im} \, G_{i}^{(N)}(z) \Big{\rangle}.
\label{rhoResolv} 
\end{equation}
The right hand side of this equation is the limit $\eta \rightarrow 0^{+}$ 
of a sum of Lorentzian functions with width $\eta$  and centered
at $E$.  The Lorentzians quantify the distances of  $\mathcal{H}_{N}$'s eigenvalues from
the energy $E$. 
As $\eta$ approaches
the mean level spacing from above our observables will display larger and larger fluctuations, since our averages will include  smaller and smaller numbers
of eigenstates.  If $\eta$ is smaller than the level spacing then one obtains results which have no physical meaning.  Accurate results for very small 
$\eta$ are obtained only if the system size $L$ and the number of samples
${\cal N}_\epsilon$ are large enough. 
Fig. \ref{sizeeffect} shows how this issue influences the IPR.  We fix the number of samples ${\cal N}_\epsilon$ and spectral line width $\eta$ 
and vary the system size.  Convergence is obtained at $L\geq 2^{26}$.
\begin{figure}[t!]
\center
\includegraphics[scale=0.9]{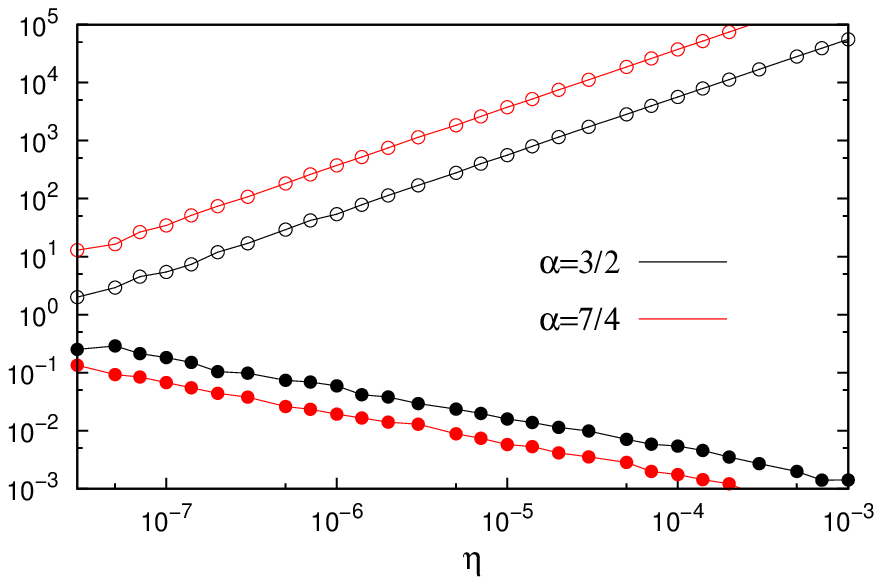}
\caption{$\eta$ dependence of the ratios $\eta/\Delta_{L,\mathcal{N},\eta}(E)$ (open circles)
and $\sigma_{L,\mathcal{N},\eta}(E)/\rho_{L,\mathcal{N},\eta}(E)$ (filled circles).  
$\Delta_{L,\mathcal{N},\eta}(E)$ is the approximate mean level spacing and $\sigma_{L,\mathcal{N},\eta}(E)$ is the standard deviation around $\rho_{L,\mathcal{N},\eta}(E)$.  Results were obtained using Eqs. (6-9) in the main text with $N=28$ and ${\cal N}_\epsilon=100$.  We set $(W=0.8,E=5.405)$ for $\alpha=3/2$, and
$(W=0.2,E=3.532)$ for $\alpha=7/4$. These values were used to produce the left-most (smallest disorder) data point in Fig. 5 of the main text.
 }
\label{etaeffect}
\end{figure}

If we define $\rho_{L,\mathcal{N},\eta}(E)$ as the average
DOS of a finite though very large system, 
we can estimate the mean level
spacing $\Delta_{L,\mathcal{N},\eta}(E)$ around $E$  as 
\begin{equation}
\Delta_{L,\mathcal{N},\eta}(E) \sim \frac{1}{ L  {\cal N}_\epsilon  
\rho_{L,\mathcal{N},\eta}(E)} .
\nonumber
\end{equation}    
We estimate the error at small $\eta$  by calculating $\sigma_{L,\mathcal{N},\eta}(E)/\rho_{L,\mathcal{N},\eta}(E)$
and $\eta/\Delta_{L,\mathcal{N},\eta}$, where $\sigma_{L,\mathcal{N},\eta}(E)$ is the standard deviation
of $\rho_{L,\mathcal{N},\eta}(E)$.   Typical results  are displayed in
Fig. \ref{etaeffect}.  When we decrease $\eta \rightarrow 0^{+}$ we find monotonic growth in  $\sigma_{L,\mathcal{N},\eta}(E)/\rho_{L,\mathcal{N},\eta}(E)$
and monotonic decay in  $\eta/\Delta_{L,\mathcal{N},\eta}$.  
For small enough $\eta$ we reach a regime where $\sigma_{L,\mathcal{N},\eta}(E)/\rho_{L,\mathcal{N},\eta}(E)=O(1)$, $\eta/\Delta_{L,\mathcal{N},\eta} = O(1)$, and   
$\rho_{L,\mathcal{N},\eta}(E)$ exhibits large fluctuations.
We conclude that the limit
$\eta \rightarrow 0^{+}$ is achieved, for practical purposes,
when $\Delta_{L,\mathcal{N},\eta}  \ll \eta \ll 1$, i.e. in very large systems. Therefore we  establish a sensible lower cutoff on $\eta$ by imposing a maximum
value of the average DOS's relative error $\sigma_{L,\mathcal{N},\eta}(E)/\rho_{L,\mathcal{N},\eta}(E)$.

The results for the minimum IPR displayed in Fig. 5 of the main text
were obtained by choosing $\eta=10^{-5}$ as the lower cutoff when 
$\alpha = 3/2$ and $\eta \in [10^{-7},10^{-5}]$ when $\alpha=7/4$.
This ensures that 
the relative error $\sigma_{L,\mathcal{N},\eta}(E)/\rho_{L,\mathcal{N},\eta}(E)$
is restricted to the interval $[10^{-2},10^{-1}]$, as can be seen in
Fig. \ref{etaeffect}.


\end{document}